# Mean-Field and Pairing Properties of Exotic Nuclei: Exploring the Nuclear Landscape

Jacek DOBACZEWSKI[1] and Witold NAZAREWICZ[1-3]

[1]*Institute of Theoretical Physics, Warsaw University, ul. Hoża 69, PL-00681, Warsaw, Poland*
[2]*Department of Physics and Astronomy, University of Tennessee, Knoxville, Tennessee 37996*
[3]*Physics Division, Oak Ridge National Laboratory, Oak Ridge, Tennessee 37831*

In years to come, we shall see substantial progress in our understanding of nuclear structure – a rich and many-faceted field. An important element in this task will be to extend the study of nuclei into new domains. The journey to "the limits" of isospin, angular momentum, and mass and charge is a quest for new and unexpected phenomena which await us in uncharted territories. What is extremely important from a theoretical point of view is that the new data are also expected to bring qualitatively new information about the effective nucleon-nucleon interaction and hence about the fundamental properties of the nucleonic many-body system. The main objective of this presentation is to discuss some of the challenges and opportunities for nuclear structure research with radioactive nuclear beams.

## §1. Introduction

The range of unstable nuclei accessible with radioactive nuclear beam (RNB) facilities opens up enormous opportunities for the study of nuclear structure, nuclear astrophysics, and fundamental interaction physics [1), 2), 3)]. Intriguing possibilities occur both at the drip lines and in the long iso-chains of nuclei between the valley of stability and the extremes of nuclear existence. Exotica in the latter region are almost sure to appear since the mean field in weakly bound neutron-rich nuclei is modified relative to nuclei near stability, since new types of correlations are likely to occur, and since reduced binding should modify residual interactions among the outermost nucleons. Between the regions of known and near-drip-line nuclei lies an extensive zone where studies will reveal much about the microscopy of the nucleus.

In this paper, we intend to discuss – rather briefly – several theoretical challenges related to the mean-field description of exotic nuclei. We focus on several recent developments and trends while the reader is referred to the review [4)] for more discussion on earlier studies and aspects not covered here.

## §2. Physics of Neutron-Rich Nuclei

One of the main avenues addressed by radioactive ion beams is the evolution of nuclear structure as a function of neutron-to-proton asymmetry.

Figure 1 shows various domains of nuclear matter, important in the context of the RNB program. The range of neutron excess, $(N-Z)/A$, in finite nuclei is from about

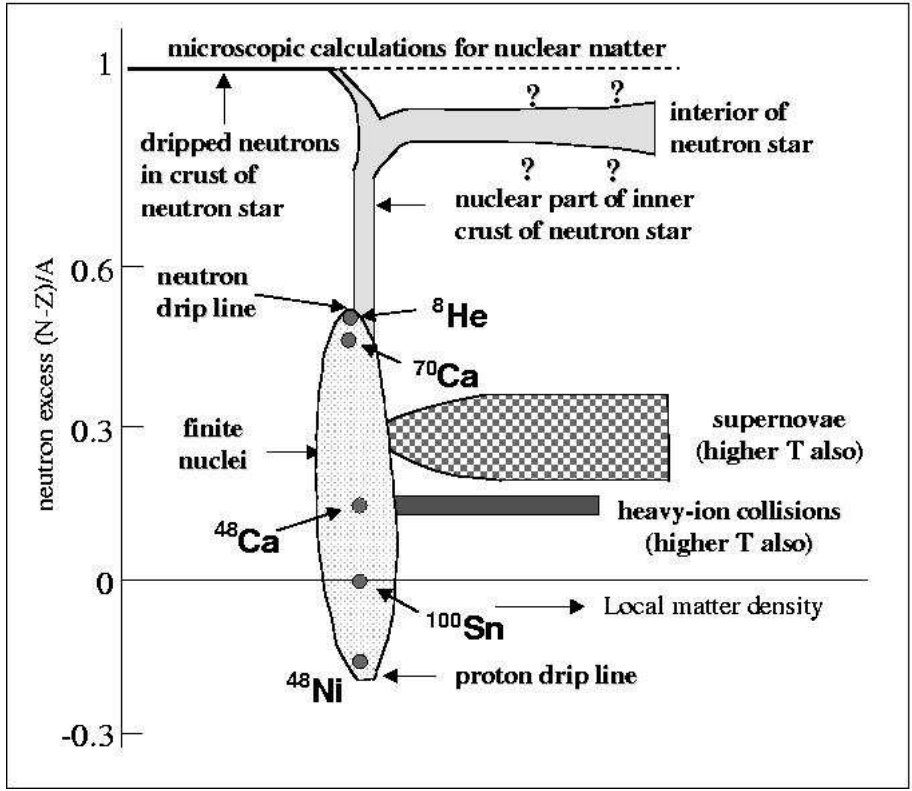

Fig. 1. Diagram illustrating the range of isoscalar densities and neutron excess, $(N-Z)/A$, important in various areas of nuclear physics and astrophysics (based on Ref.[5]).

–0.2 (proton drip line) to 0.5 (neutron drip line). Of course, by using radioactive beams (and radioactive targets), one will be able to explore new regions of Fig. 1. In particular, the new-generation RNB facilities, such as the Rare Isotope Accelerator, will provide a unique capability for accessing the very asymmetric nuclear matter and for compressing neutron-rich matter approaching density regimes important for supernova and neutron star physics.

From a theoretical point of view, exotic neutron-rich nuclei far from stability are of particular interest. They offer a unique test of those components of effective interactions that depend on the isospin degrees of freedom. In many respects, weakly bound nuclei are much more difficult to treat theoretically than well-bound systems[4]. For weakly bound nuclei, the Fermi energy lies very close to zero, and the particle continuum must be taken into account explicitly.

2.1. *Uncertain Extrapolations*

Experimentation with radioactive nuclear beams is expected to expand the range of known nuclei. That is, by going to nuclei with extreme $N/Z$ ratios, one can magnify the isospin-dependent terms of the effective interaction (which are small in "normal" nuclei). The hope is that after probing these terms at the limits of extreme isospin, we can later go back to the valley of stability and improve the description of normal nuclei. In addition to nuclear structure interest, the understanding of effective interactions in the neutron-rich and proton-rich environment is important for astrophysics and cosmology.

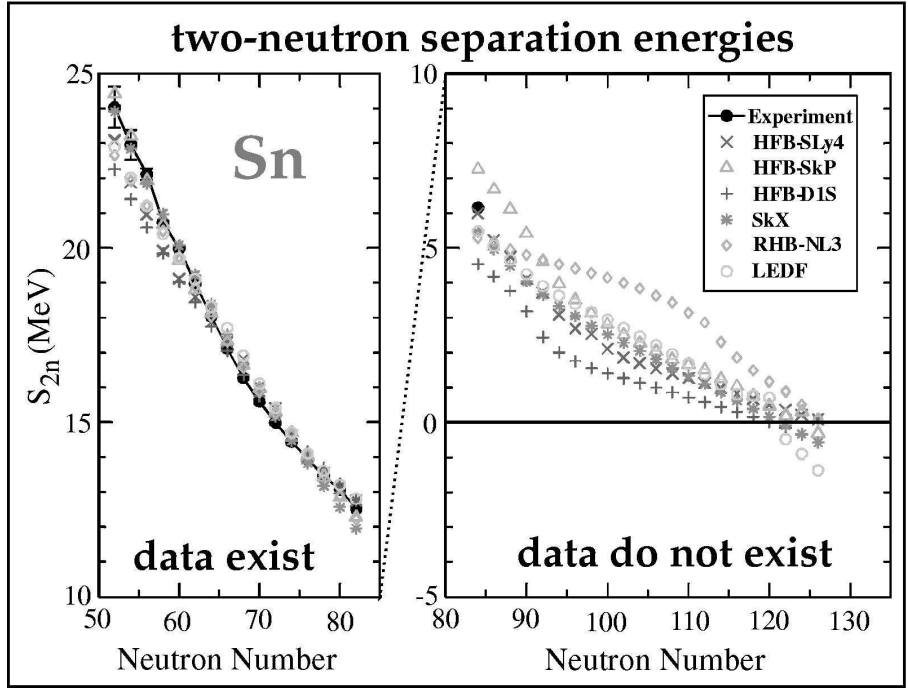

Fig. 2. Predicted two-neutron separation energies for the even-even Sn isotopes using several microscopic models based on effective density-dependent nucleon-nucleon interactions (from Refs. [6], [7]).

Figure 2 illustrates difficulties with making theoretical extrapolations into neutron-rich territory. It shows the two-neutron separation energies for the even-even Sn isotopes calculated in several microscopic models based on different effective interactions. Clearly, the differences between forces are greater in the region of "terra incognita" than in the region where masses are known (an effect half-jokingly called "asymptotic freedom"). Therefore, the uncertainty due to the largely unknown isospin dependence of the effective force (in both particle-hole and particle-particle channels) gives an appreciable theoretical "error bar" for the position of the drip line. Unfortunately, the results presented in Fig. 2 do not tell us much about which of the forces discussed should be preferred since one is dealing with dramatic extrapolations far beyond the region known experimentally. However, a detailed analysis of the force dependence of results may give us valuable information on the relative importance of various force parameters. Moreover, it is obvious that the experimental data on nuclei near stability do not sufficiently constrain results for exotic nuclei. Therefore, further measurements in neutron-rich systems are of vital importance for the development of theoretical descriptions. We note in passing that attempts to derive the effective interactions from first principles may lead to even

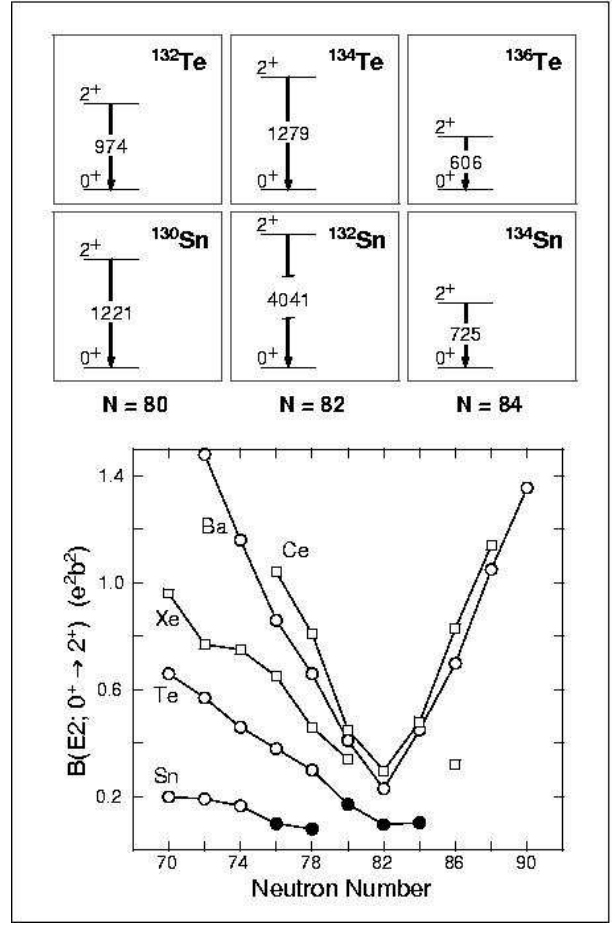

Fig. 3. Top: $2_1^+$ levels in $N$=80,82,84 Sn and Te isotopes. Bottom: Values of $B(E2; 0^+ \to 2_1^+)$ for even-even Sn, Te, Xe, Ba, and Ce isotopes around neutron number $N$=82. Filled symbols indicate the recent RNB measurements at the HRIBF facility at ORNL (from Ref. [9]).

larger uncertainties for exotic nuclei (see recent analysis in Ref. [8]).

Figure 3 demonstrates that one does not need to go all the way to the neutron drip line to see deviations from well-established trends. Indeed, the differences between various models describing masses of Sn isotopes show up just above $N$=82. Interestingly, experimental data around $^{132}$Sn exhibit an unusual pattern as one crosses $N$=82. Namely, there is a striking asymmetry in the position of $2_1^+$ levels in $N$=80 and 82 isotopes of Sn and Te, and the $B(E2; 0^+ \to 2_1^+)$ rate in $^{136}$Te stays unexpectedly low [9], defying common wisdom that the decrease in $E_{2_1^+}$ in open-shell nuclei must imply the increase in $B(E2; 0^+ \to 2_1^+)$. So far, there is no satisfactory explanation for the pattern shown in Fig. 3. What is clear, however, is that one must be prepared for many surprises when entering the neutron-rich territory.

### 2.2. *Isospin Dependence of Pairing*

The uniqueness of drip-line nuclei for studies of effective interactions is due to the very special role played by the pairing force. Correlations due to pairing, core polarization, and clustering are crucial in weakly bound nuclei. In a drip-line system, the pairing interaction and the presence of skin excitations (soft modes) could invalidate the picture of a nucleon moving in a single-particle orbit [10,11,12,13,14].

Surprisingly, rather little is known about the basic properties of the pairing force. In most nuclear structure calculations, the pairing Hamiltonian has been approximated by the state-independent seniority pairing force, or schematic multipole pairing interaction [15]. Such oversimplified forces, usually treated by means of the BCS approximation, perform remarkably well when applied to nuclei in the neighborhood of the stability valley, but they are inappropriate (and formally wrong) when extrapolating far from stability. The self-consistent mean-field models have meanwhile reached such a high level of precision that one needed to improve on the pairing part of the model. Presently, the most up-to-date models employ local pairing forces parametrized as contact interactions [16,17,10]. More flexible forms attach a density-dependence to the pairing strength [18,19,20]. There exist even more elaborate forms which include also gradient terms [21,22]. Although all these various density dependencies of pairing are widely used, very little is yet known about their relations to observable quantities.

Up to now, the microscopic theory of the pairing interaction has only seldom been applied in realistic calculations for finite nuclei. A "first-principle" derivation of pairing interaction from the bare $NN$ force still encounters many problems such as, e.g., treatment of core polarization [23,24]. Hence, phenomenological density-dependent pairing interactions are usually introduced. It is not obvious, how the density dependence should be parametrized [10], although nuclear matter calculations and experimental data on isotope shifts strongly suggest that pairing is a surface phenomenon, and that pairing interaction should be maximal in the surface region. This is why neutron-rich nuclei play such an important role in this discussion. Indeed, because of strong surface effects, the properties of these nuclei are sensitive to the density dependence of pairing.

Recent work [25], based on the spherical Skyrme-HFB model, contains the theoretical analysis of particle and pairing densities in neutron-rich nuclei and their dependence on the choice of pairing interaction. In the particle-particle (p-p) channel, the density-dependent delta interaction (DDDI) [18,19,20] has been employed:

$$V_{\text{pair}}^{\delta\rho}(\mathbf{r},\mathbf{r}') = f_{\text{pair}}(\mathbf{r})\delta(\mathbf{r}-\mathbf{r}'), \tag{2.1}$$

where the pairing-strength factor is

$$f_{\text{pair}}(\mathbf{r}) = V_0\left\{1 - [\rho_{\text{IS}}(\mathbf{r})/\rho_c]^\alpha\right\} \tag{2.2}$$

and $V_0$, $\rho_c$, and $\alpha$ are constants. The presence of the density dependence in the pairing channel has consequences for the spatial properties of pairing densities and fields [16,17,10]. In Eq. (2.2) $\rho_{\text{IS}}(\mathbf{r})$ stands for the isoscalar single-particle density $\rho_{\text{IS}}(\mathbf{r})=\rho_n(\mathbf{r})+\rho_p(\mathbf{r})$. If $\rho_c$ is chosen such that it is close to the saturation density, $\rho_c \approx \rho_{\text{IS}}(\mathbf{r}=0)$, both the resulting pair density and the pairing potential $\tilde{h}(\mathbf{r})$ are small in the nuclear interior, and the pairing field becomes surface-peaked. By varying the magnitude of the density-dependent term, the transition from volume pairing to surface pairing can be probed.

Apart from rendering the pairing weak in the interior, the specific functional dependence on $\rho_{\text{IS}}$ used in Eq. (2·2) is not motivated by any compelling theoretical arguments or calculations. In particular, values of power $\alpha$ were chosen *ad hoc* to be either equal to 1 (based on simplicity), see, e.g., Refs. [26), 27)], or equal to the power $\gamma$ of the Skyrme-force density dependence in the p-h channel [17), 10)]. As a typical example, the particle and pairing local HFB+SLy4 neutron densities $\rho_n(r)$ and $\tilde{\rho}_n(r)$ calculated for several values of $\alpha$ are displayed in Fig. 4 for $^{150}$Sn.

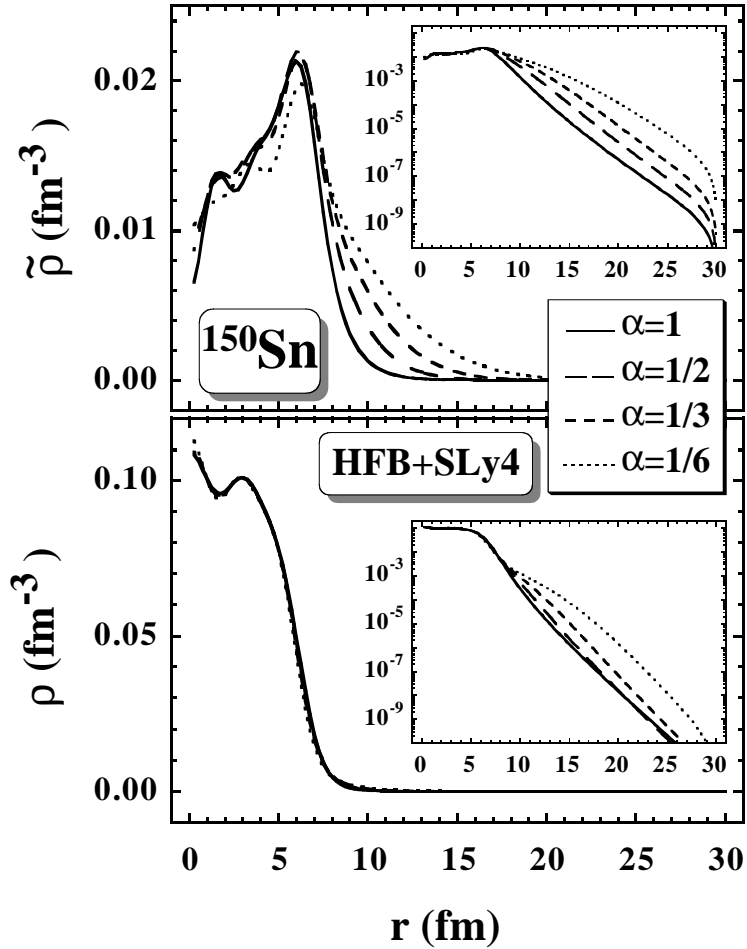

Fig. 4. Self-consistent spherical HFB+SLy4 local densities $\rho(r)$ (top) and $\tilde{\rho}(r)$ (bottom) for neutrons in $^{150}$Sn for several values of $\alpha$. The insets show the same data in logarithmic scale.

With decreasing $\alpha$, the p-p density $\tilde{\rho}_n(r)$ develops a long tail extending towards large distances. This is a direct consequence of the attractiveness of DDDI at low densities when $\alpha$ is small. While in the nuclear interior, the p-h density $\rho_n(r)$ depends extremely weakly on the actual form of pairing interaction. Due to the self-consistent feedback between particle and pairing densities, the asymptotic values of $\rho_n(r)$ are significantly increased when $\alpha$ gets small (see inset). Moreover, one observes a clear development of a halo structure, i.e., a smooth exponential decrease, that for $\alpha=1$ starts at $r\simeq 6$ fm, is interrupted at $r\simeq 9$ fm for small $\alpha$, and replaced by a significantly slower decrease of the density. The general conclusion drawn from Fig. 4 is that experimental studies of neutron distributions in nuclei are extremely important for determining the density dependence of pairing interaction in nuclei.

While the analysis presented in Ref. [25] strongly suggested that the strong low-density dependence of pairing force, simulated by taking very small values of $\alpha$ in DDDI, is unphysical, it is only very recently that a realistic fit of DDDI to the odd-even staggering in nuclear masses has been carried out [28].

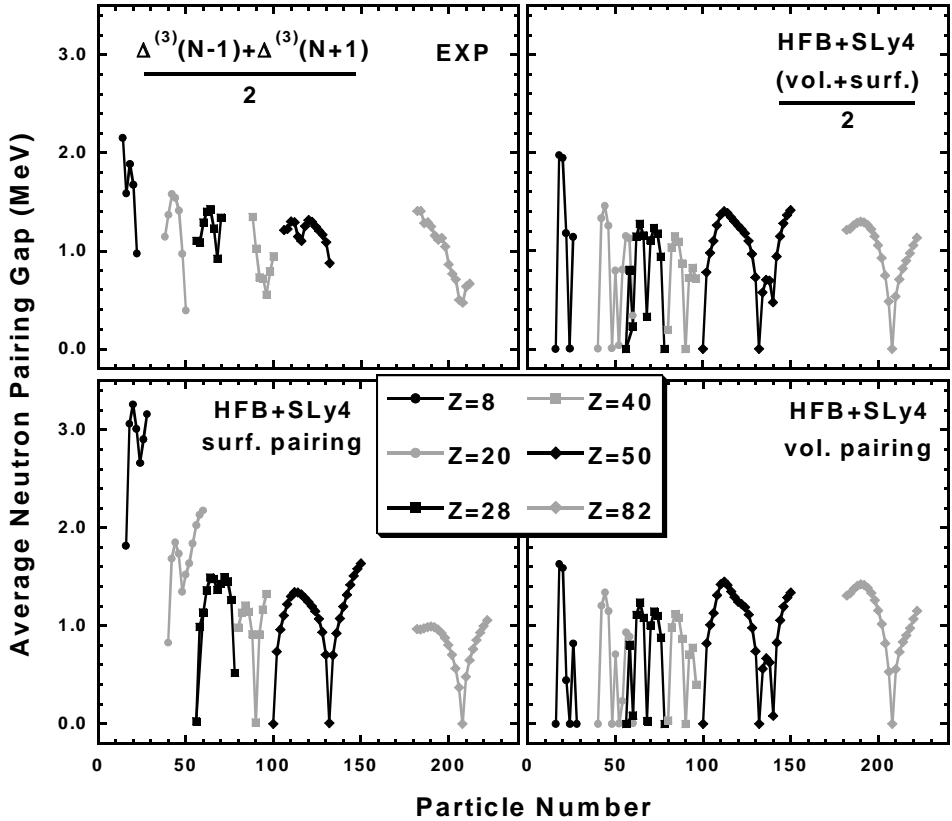

Fig. 5. Comparison between the experimental staggering parameters (upper left panel, based on masses from Ref. [29]) and the average neutron pairing gaps calculated within the spherical HFB method for the Skyrme SLy4 force and three different versions of the zero-range pairing interaction.

Results of the spherical coordinate-space HFB calculations for semi-magic even-even nuclei for the average pairing gaps are shown in Figs. 5 and 6 for neutrons and protons, respectively. In the upper left panels we show the values of experimental three-point staggering parameters $\Delta^{(3)}$ centered at odd particle numbers [30), 31)] and averaged over the two particle numbers adjacent to the even value. The experimental data from the interim 2001 atomic mass evaluation [29)] were used.

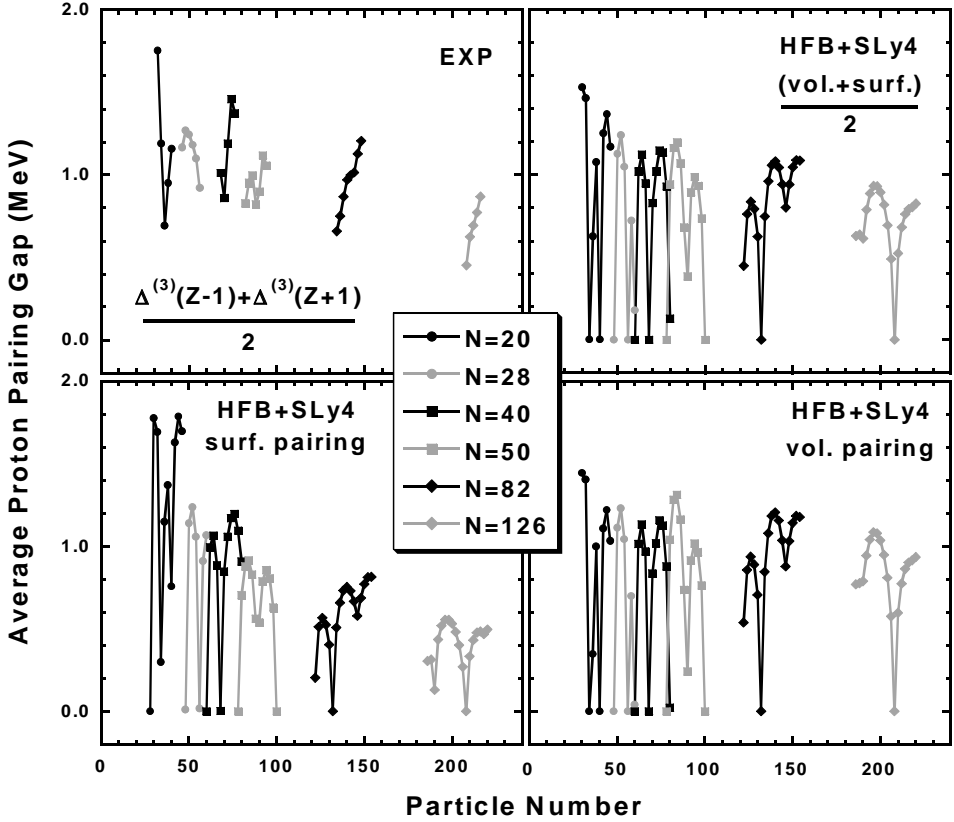

Fig. 6. Same as in Fig. 5 except for the average proton pairing gaps.

The lower left and right panels in Figs. 5–6 show the results obtained for the surface ($\alpha=1$) and volume pairing interactions, respectively. When compared with the experimental numbers, one sees that both types of pairing interaction are in clear disagreement with experiment. The surface interaction gives the pairing gaps that increase very rapidly in light nuclei, while the volume force gives the values that are almost independent of $A$. The surface pairing in light nuclei is so strong that pairing correlations do not vanish in doubly magic nuclei such as $^{16}$O or $^{40}$Ca. The experimental data show the trend that is intermediate between surface and volume; hence, one may consider the intermediate-character pairing force that is half way in between, i.e., it is defined as:

$$V^\delta_{\rm mix}(\mathbf{r},\mathbf{r}') = \frac{1}{2}\left(V^\delta_{\rm vol} + V^\delta_{\rm surf}\right) = V_0\left[1 - \frac{\rho(\mathbf{r})}{2\rho_0}\right]\delta(\mathbf{r}-\mathbf{r}'). \qquad (2\cdot 3)$$

The upper right panels in Figs. 5 and 6 show the results obtained with the mixed pairing force. It can be seen that one obtains significantly improved agreement with the data, although a more precise determination of the balance between the surface and volume contributions still seems to be possible. One should note that similar intermediate-character pairing forces have recently been studied in Ref. [32)] where it was concluded that pairing in heavy nuclei is of a mixed nature.

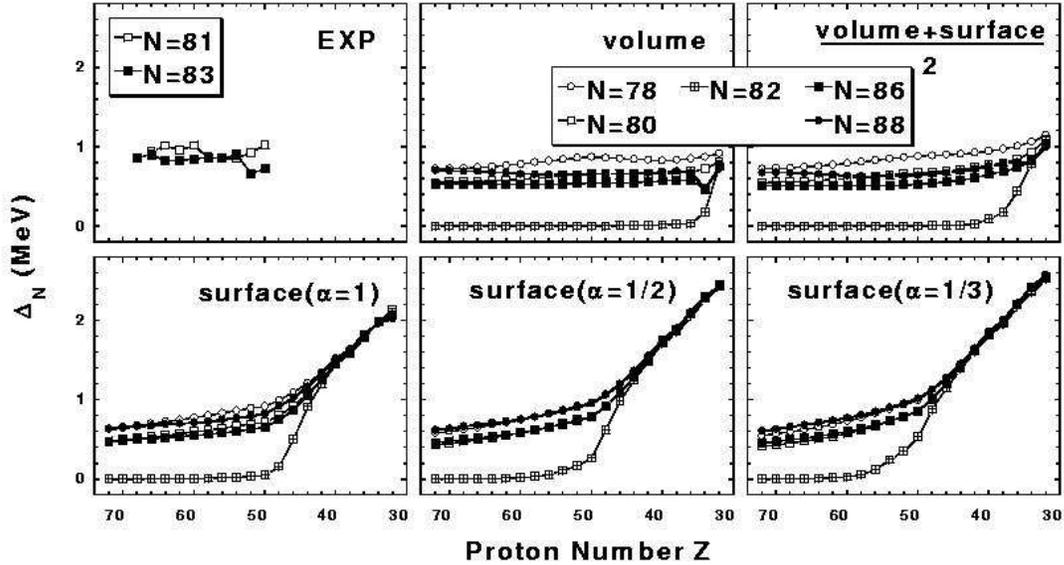

Fig. 7. Similar as in Fig. 5, but for chains of isotones around $N=82$. The experimental neutron pairing gaps were estimated from the three-point [30), 31)] staggering parameters $\Delta^{(3)}$ centered at $N=81$ and $N=83$ (from Ref. [33)]).

Figure 7 illustrates the role of using different types of the pairing interaction to predict the neutron pairing gaps in very neutron-rich nuclei. The experimental data that exist for $Z \geq 50$ do not indicate any definite change in the neutron pairing intensity with varying proton numbers. However, the surface pairing interactions (bottom panels) give a slow dependence for $Z \geq 50$ that is dramatically accelerated after crossing the shell gap at $Z=50$. On the other hand, the volume and mixed pairing forces predict a slow dependence all the way through to very near the neutron drip line. It is clear that measurements of only several nuclear masses for $Z<50$ will allow us to strongly discriminate between the pairing interactions that have different space and density dependencies.

## §3. Physics of Very Heavy Nuclei

The stability of the heaviest and superheavy elements has been a longstanding fundamental question in nuclear science. Theoretically, the mere existence of the heaviest elements with $Z>102$ is entirely due to quantal shell effects. Indeed, for these nuclei the shape of the classical nuclear droplet, governed by surface tension and Coulomb repulsion, is unstable to surface distortions driving these nuclei to spontaneous fission. That is, if the heaviest nuclei were governed by the classical liquid drop model, they would fission immediately from their ground states due to the large electric charge. However, in the mid-sixties, with the invention of the shell-correction method, it was realized that long-lived superheavy elements (SHE) with very large atomic numbers could exist due to the strong shell stabilization [34), 35)].

The last three years (1999-2001) have brought a number of experimental surprises which have truly rejuvenated the field (see Ref. [36)] for a recent review). Most significant are reports from Dubna announcing the discovery of elements 114 and 116 in hot fusion reactions [37), 38), 39)]. Further experiments with stable beams are planned; they will be extremely helpful for the theoretical modeling of the SHE formation. It is anticipated, however, that future experimental progress in the synthesis of new elements will be possible – thanks to radioactive nuclear beams, especially the doubly magic neutron-rich beam of $^{132}$Sn [36)].

In spite of an impressive agreement with experimental data for the heaviest elements, theoretical uncertainties are large when extrapolating to unknown regions of the nuclear chart. In particular, there is no consensus among theorists with regard to the center of the shell stability in the superheavy region. Since in these nuclei the single-particle level density is relatively large, small shifts in the position of single-particle levels (e.g., due to the Coulomb or spin-orbit interaction) can be crucial for determining the shell stability of a nucleus. But does it actually make sense to talk about "magic superheavy nuclei"? Recent theoretical work [40)] sheds new light on this question. According to calculations, the patterns of single-particle levels are significantly modified in the superheavy elements. Firstly, the overall level density grows with mass number $A$, as $\propto A^{1/3}$. Secondly, no pronounced and uniquely preferred energy gaps appear in the spectrum. This shows that shell closures which are to be associated with large gaps in the spectrum are not robust in superheavy nuclei. Indeed, the theory predicts that beyond $Z = 82$ and $N = 126$ the usual localization of shell effects at magic numbers is basically gone. Instead the theory predicts fairly wide areas of large shell stabilization without magic gaps (see Fig. 8). This is good news for experimentalists: there is a good chance to reach shell-stabilized superheavy nuclei using a range of beam-target combinations.

The Coulomb and nuclear interactions act in opposite ways on the total nucleonic density in the nuclear interior. As a consequence of their saturation properties, nuclear forces favor values of the internal density close to the saturation density of nuclear matter. On the contrary, since the Coulomb interaction tends to increase the average distance between protons, the Coulomb energy is significantly lowered by either the creation of a central depression or by deformation, or both. Based on this general argument, one expects the formation of voids in heavy nuclei in which the Coulomb energy is very large [41), 42)]. Recently, the subject of exotic (bubble, toroidal, band-like) configurations in nuclei with very large atomic numbers has been revisited by self-consistent calculations. The important question which is asked in this context is: What are the properties of the heaviest nuclei that can be bound (at least, in theory), in spite of the tremendous Coulomb force?

Calculations do predict the existence of bubble nuclei [43), 44), 45)] which are stabilized by shell effects [46)]. Figure 9 shows the results of the coordinate-space SLy6+HF calculations with delta pairing [47)]. The potential energy surfaces (PES) corresponding to various density distributions are plotted as functions of the mass quadrupole deformation $\beta_2$. This figure nicely illustrates the competition between various forms of nuclear matter constituting a finite heavy nucleus possessing a huge electric charge. It is difficult to say at present whether these exotic topologies can occur as metastable states or whether they can form isolated islands of nuclear stability stabilized by shell effects. The interplay between exotic configurations in superheavy nuclei is likely to impact the exact position of the borders of the superheavy and hyperheavy region.

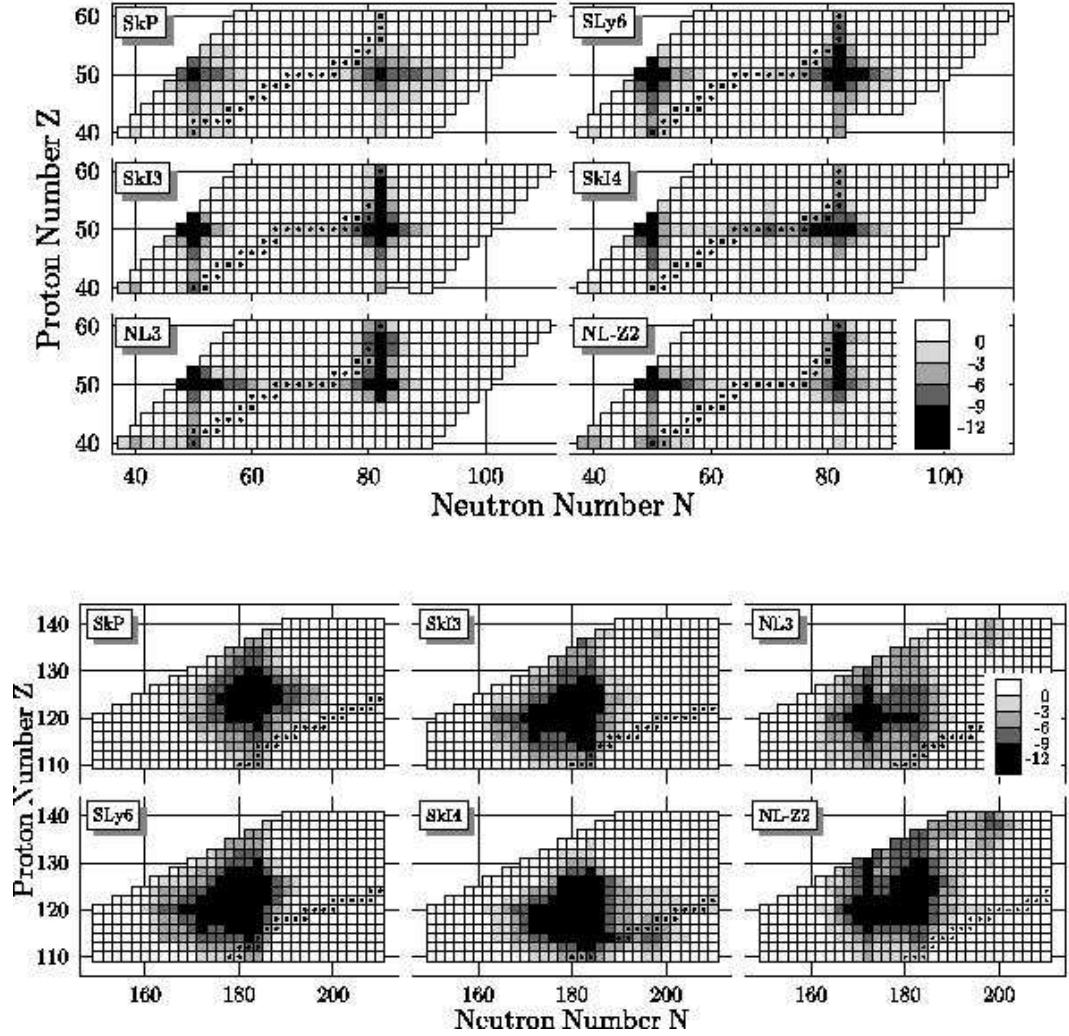

Fig. 8. Total shell energy (sum of proton and neutron shell corrections) calculated for spherical even-even nuclei with $40 \leq Z \leq 60$ (top) and for superheavy nuclei around the expected island of stability around $Z=120$, $N=180$. Dots mark nuclei predicted to be stable with respect to $\beta$-decay (from Ref. [40]).

## §4. Conclusions

The main objective of this brief review was to discuss various facets of RNB physics. The list of topics covered is, of course, very limited due to time and space constraints. An experimental excursion into uncharted territories of the chart of the nuclides, exploring new combinations of $Z$ and $N$, will offer many excellent opportunities for nuclear structure research. What is most exciting, however, is that there are many unique features of exotic nuclei that give prospects for entirely new phenomena likely to be different from anything we have observed to date. We are only at the beginning of this most exciting journey.

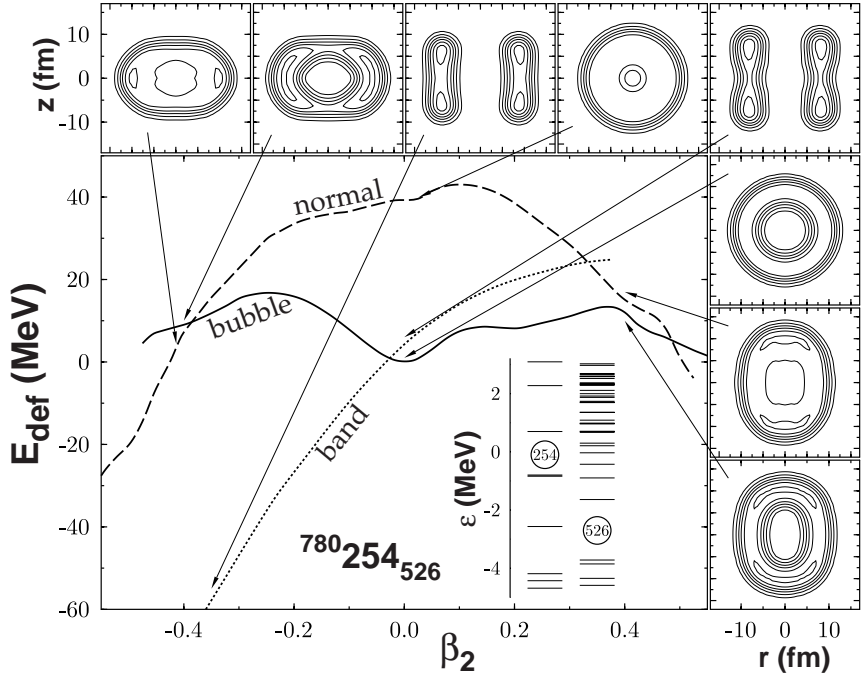

Fig. 9. Variety of shapes predicted for the superheavy nucleus $^{780}254_{526}$ in the SLy6+HF model with delta pairing. Axial symmetry was assumed. The local minimum at $\beta_2=0$ can be associated with a bubble configuration (solid line). At small deformations, the PES corresponding to the "normal" density distribution (dashed line) is predicted to lie very high in energy. At small oblate deformations, the unstable "band"-like structure becomes favored energetically. The single-particle levels corresponding to the spherical bubble minimum are shown in the inset. The contour plots of the corresponding total densities are given in the boxes. The contours are axially symmetric with respect to $r=0$ and the contour lines are at 0.01, 0.03, 0.06, 0.09, 0.12, and 0.15 fm$^{-3}$ (from Ref. [47]).

## Acknowledgments

This research was supported in part by the U.S. Department of Energy under Contract Nos. DE-FG02-96ER40963 (University of Tennessee) and DE-AC05-00OR22725 with UT-Battelle, LLC (Oak Ridge National Laboratory), by the Polish Committee for Scientific Research (KBN) under Contract No. 5 P03B 014 21, and by the computational grant from the Interdisciplinary Centre for Mathematical and Computational Modeling (ICM) of the Warsaw University.